\documentclass[12pt]{article}
\usepackage{graphicx}
\usepackage{color}
\usepackage{cite}
\usepackage{url}

\newcommand{\gsim}{\lower.7ex\hbox{$\;\stackrel{\textstyle>}{\sim}\;$}}
\newcommand{\lsim}{\lower.7ex\hbox{$\;\stackrel{\textstyle<}{\sim}\;$}}

\def \beq{\begin{equation}}
\def \eeq{\end{equation}}
\def\eqref#1{(\ref{#1})}
\def\bea{\begin{eqnarray}}
\def\eea{\end{eqnarray}}
\def\jpsi{\hbox{$J\kern-0.2em/\kern-0.1em\psi$}}
\def\Y1S{\hbox{$\Upsilon(1S)$}}
\def \ket#1{|{#1}\rangle}
\def \bra#1{\langle{#1}|}

\def\URLtilde{\lower0.2em\hbox{$\tilde{\phantom{a}}$}}
\def\mycomm#1{\hfill\break\strut\kern-3em{\color{red}\tt ====> #1
\color{black}}\hfill\break}

\newcount\timecount
\newcount\hours \newcount\minutes  \newcount\temp \newcount\pmhours
\hours = \time
\divide\hours by 60
\temp = \hours
\multiply\temp by 60
\minutes = \time
\advance\minutes by -\temp
\def\hour{\the\hours}
\def\minute{\ifnum\minutes<10 0\the\minutes
\else\the\minutes\fi}
\def\clock{
\ifnum\hours=0 12:\minute\ AM
\else\ifnum\hours<12 \hour:\minute\ AM
\else\ifnum\hours=12 12:\minute\ PM
\else\ifnum\hours>12
\pmhours=\hours
\advance\pmhours by -12
\the\pmhours:\minute\ PM
\fi
\fi
\fi
\fi
}

\def\monthname{\relax\ifcase\month 0/\or January\or February\or
March\or April\or May\or June\or July\or August\or September\or
October\or November\or December\else\number\month/\fi}

\def\bold#1{\setbox0=\hbox{$#1$}     \kern-.025em\copy0\kern-\wd0
\kern.05em\copy0\kern-\wd0
\kern-.025em\raise.0433em\box0 }

\begin{document}
\setcounter{footnote}{1}
\rightline{EFI 15-25}
\rightline{TAUP 2998/15}
\rightline{arXiv:1508.01496}

\begin{center}
{\Large \bf Photoproduction of Exotic Baryon Resonances}
\end{center}
\bigskip

\centerline{\bf Marek Karliner$^a$\footnote{{\tt marek@proton.tau.ac.il}}
 and Jonathan L. Rosner$^b$\footnote{{\tt rosner@hep.uchicago.edu}}}
\medskip

\centerline{$^a$ {\it School of Physics and Astronomy}}
\centerline{\it Raymond and Beverly Sackler Faculty of Exact Sciences}
\centerline{\it Tel Aviv University, Tel Aviv 69978, Israel}
\medskip

\centerline{$^b$ {\it Enrico Fermi Institute and Department of Physics}}
\centerline{\it University of Chicago, 5620 S. Ellis Avenue, Chicago, IL
60637, USA}
\bigskip
\strut

\begin{center}
ABSTRACT
\end{center}
\begin{quote}
We point out that the new exotic resonances recently reported by LHCb in the
$\jpsi\,p$ channel are excellent candidates for photoproduction off a proton
target.   This test is crucial to confirming the resonant nature of such
states, as opposed to their being kinematical effects.  We specialize to an
interpretation of the heavier narrow state as a
molecule composed of $\Sigma_c$ and $\bar D^*$, and estimate its production
cross section using vector dominance.  The relevant photon energies and fluxes
are well within the capabilities of the GlueX and CLAS12 detectors at Thomas
Jefferson National Accelerator Facility (JLAB).  A corresponding calculation is
also performed for photoproduction of an analogous resonance which is predicted
to exist in the $\Upsilon p$ channel.
\end{quote}

\smallskip

\leftline{PACS codes: 12.39.Hg, 12.39.Jh, 14.20.Pt, 14.40.Rt}
\bigskip


\section{Introduction \label{sec:intro}}

The LHCb experiment \cite{LHCb} has observed two new exotic resonances in
the $\jpsi\,p$ channel, a broad one with mass $4380\pm8\pm29$ MeV, width
$205\pm18\pm86$ MeV, and statistical significance $9\sigma$, and a narrower
one with mass $4449.8\pm1.7\pm2.5$ MeV, width $39\pm5\pm19$ MeV, and
statistical significance $12 \sigma$.  In the present note we point out
that these states are excellent candidates for photoproduction off a proton
target, an observation made by others \cite{Wang:2015jsa,Kubarovsky:2015aaa}
as a preliminary version of this Letter was being prepared.  Specializing to
an interpretation in which the heavier state is regarded as a molecule of
$\Sigma_c$ and $\bar D^*$ \cite{mol}, we estimate the cross section for its
production using vector dominance.  A corresponding calculation is also
performed for a molecule of $\Sigma_b$ and $B^*$ forming an $\Upsilon p$
resonance.  Observation of the states observed by LHCb in photoproduction is
crucial to their confirmation as resonances as opposed to their being kinematic
enhancements.

\section{\boldmath
The reaction $\gamma \,p \to X \to \jpsi\,p$ \label{sec:jpsi}}

We calculate the cross section for photoproduction of a resonance $X$
decaying to $\jpsi\,p$ by assuming it is dominated by the elastic process
$\jpsi\,p \to X \to \jpsi\,p$.  The photon-$\jpsi$ coupling is estimated
from the $\jpsi$ leptonic width: $\Gamma(\jpsi \to \ell^+ \ell^-) = 5.55 \pm
0.14 \pm 0.02$ keV \cite{PDG}.  The Breit-Wigner cross section for production
of a resonance with spin $J$ by particles of spins $S_1$ and $S_2$ 
is \cite{PDG} 
\beq
\sigma_{BW}(E) = \frac{2J+1}{(2S_1+1)(2S_2+1)}~\frac{4\pi}{k^2_{\rm in}}~
  \frac{B_{\rm in}B_{\rm out} (\Gamma_{\rm tot}^2/4)}{(E-E_R)^2 + 
  (\Gamma_{\rm tot}^2/4)}~,
\eeq
where $k_{\rm in,out}$ are the center-of-mass (CM) 3-momenta in the (incoming
$\gamma p$, outgoing $J/\psi~p)$ channel, $E=E_{\rm cm}$ is the total CM
energy, $E_R$ is the resonance energy, $B_{\rm in}$ and $B_{\rm out}$ are the
resonance branching fractions into the incoming and outgoing channels, and
$\Gamma_{\rm tot}$ is the resonance total width.  For $E_R = 4380$ MeV,
$k^A_{\rm in,out} = (2090,741)$ MeV (we use units in which $c=1$), while for
$E_R = 4450$ MeV, $k^B_{\rm in,out} = (2126,820)$ MeV.  (We shall denote these
resonances $X_A$ and $X_B$, respectively.)  In the preferred fits of Ref.\
\cite{LHCb}, one of these resonances has spin 3/2, the other has spin 5/2, and
they are of opposite parity.  One theoretical interpretation of the narrow
higher-lying state as a $\Sigma_c \bar D^*$ molecule bound by pion exchange
\cite{mol} assigns its spin and parity to be $J_B^P = 3/2^-$ and therefore
$J_A^P=5/2^+$.  For an incident photon, with only transverse polarizations,
the $2S_1+1$ factor in the denominator is to be multiplied by 2/3.

We define the decay constant $f_V$ of a vector meson $V$ in terms of the matrix
element between the one-$V$ state and the vacuum:
\beq
\bra{0}V_\mu \ket{V(q,\epsilon_\mu)} = \epsilon_\mu M_V f_V~,
\eeq
where $q$, $\epsilon$, and $M_V$ are the four-momentum, polarization
vector, and mass of the vector meson.  Then dominance of the photoproduction
cross section by the $\jpsi$ pole implies\footnote{We thank M. Voloshin for
a correction to a preliminary version of this Letter.}
\beq
B_{\rm in}/B_{\rm out}=(e f_{\jpsi} /M_{\jpsi})^2 f_L (k_{\rm in}/k_{\rm out})
^{2L+1}~,
\eeq
where $f_L$ is the fraction of decays $P_c \to J/\psi~p$ in a relative partial
wave $L$ that give rise to a transversely polarized $J/\psi$.  With our $J^P$
assignments, $L=1,3$ for $X_A = P_c(4380)$ and $L=0,2$ for $X_B = P_c(4450)$.

The leptonic width of the $\jpsi$ (neglecting lepton masses) is
\beq
\Gamma(\jpsi \to \ell^+ \ell^-) = \frac{4 \pi \alpha^2}{3}~\frac{f_{\jpsi}^2}
  {M_{\jpsi}}~,
\eeq
from which, using the experimental central value \cite{PDG}, we find
\beq
f_{\jpsi} = 278~{\rm MeV}~,~~B_{\rm in}/B_{\rm out} = 7.37 \times 10^{-4}
f_L (k_{\rm in}/k_{\rm out})^{2L+1}~.
\eeq

For subsequent purposes we shall consider only the photoproduction of the
state $X_B$ decaying to $J/\psi~p$ with relative orbital angular momentum
$L=0$, so henceforth $f_L \equiv f_0$.  It may be easily seen that the cases
$L=2$ for $X_B$ and $L=1,3$ for
$X_A$ production lead to higher predicted cross sections, so our estimate may
be regarded as a lower bound.  The quantity $f_0$ is given by $f_0 = 2/(2 +
\gamma^2) = 0.651$, where $\gamma^2 = 1 + (k^B_{\rm out}/M_{J/\psi})^2 = 1.070$
accounts for the relativistic enhancement of the longitudinally polarized
$J/\psi$ degree of freedom.  This leads to $B_{\rm in}/B_{\rm out} = 1.24
\times 10^{-3}$.  Then the cross section for $X_B$ production is
\beq \label{eqn:jpsip}
\sigma_{BW}(E) = \frac{C_B (B_{\rm out})^2 (k^B_{\rm in}/k_{\rm in})^2
(\Gamma_{\rm tot}^2/4)} {(E-E_R)^2 + (\Gamma_{\rm tot}^2/4)}~,
\eeq
where $k_{\rm in} = (E^2 - m_p^2)/(2E)$ is the magnitude of the incoming
3-momentum in the CM.  For a photon on a proton target ($S_2 = 1/2$), with
$J_B = 3/2$, one has
\beq
C_B \equiv \frac{4\pi}{(k^B_{\rm in})^2} \, \ \frac{B_{\rm in}}{B_{\rm out}}\,,
\eeq
yielding $C_B = 1.35~\mu{\rm b}$.
This is a substantial cross section, considering that the diffractive
cross section for $\gamma p \to \jpsi\,p$ is below 1~nb at $E = 4.4$ GeV
\cite{Camerini:1975cy,Gittelman:1975ix,Brodsky:2000zc,Chudakov2008,%
Aid:1996dn}. We will return to this subject at the end of the current Section.

\begin{figure}[t]
\begin{center}
\includegraphics[width=1.0\textwidth]{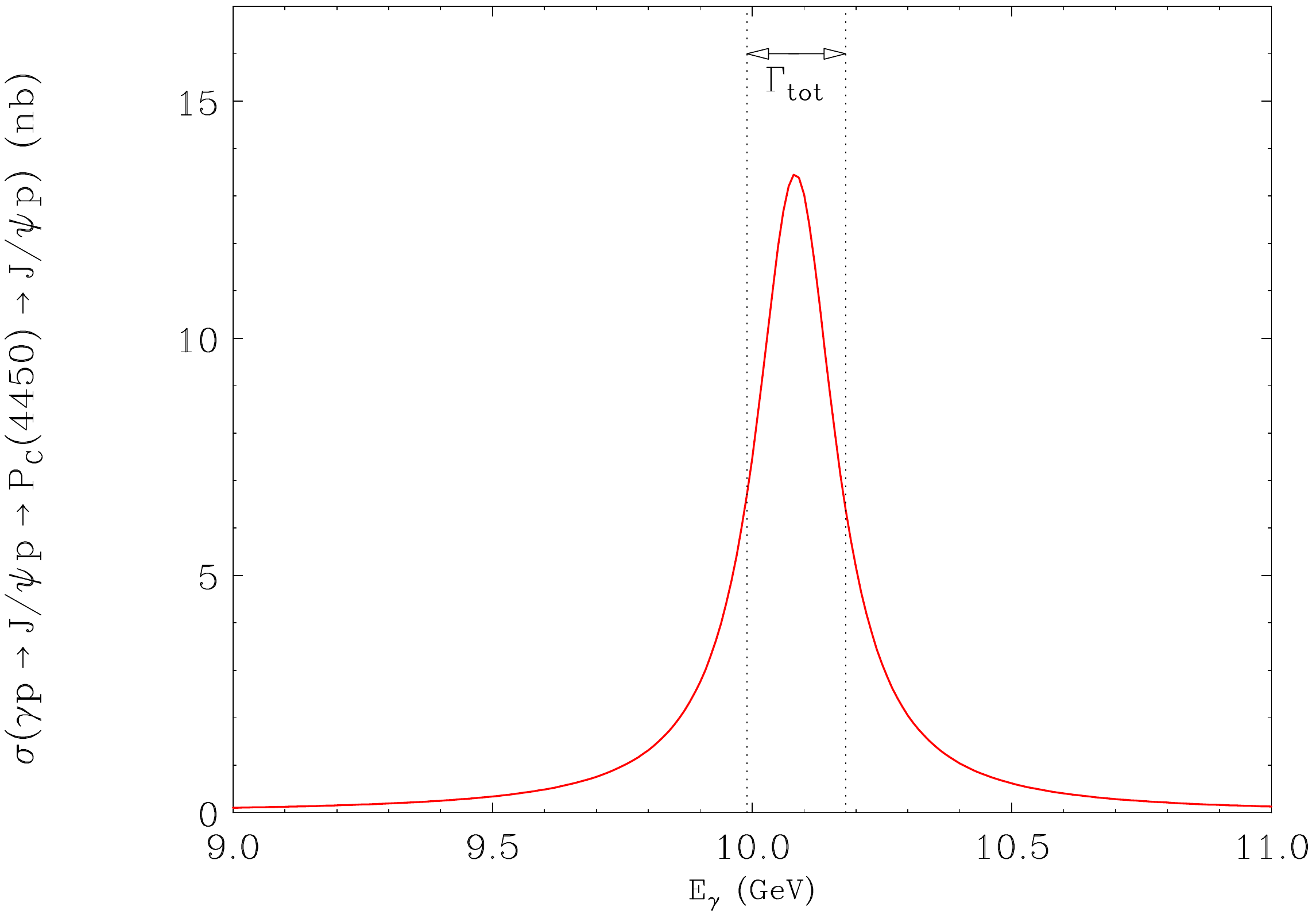}
\end{center}
\strut\vskip-1.2cm 
\caption{Cross section for resonant photoproduction $\gamma p \to \jpsi\,p \to
P_c(4450) \to \jpsi\,p$, assuming $B_{\rm out} = 0.1$, plotted as function
of the incident photon energy $E_\gamma$. The vertical dotted lines 
indicate the width of the $P_c(4450)$ resonance.}
\label{fig:sigma_photo}
\end{figure}

The size of the resonant cross sections is illustrated 
by Fig.~\ref{fig:sigma_photo} 
which shows the cross section for case (B), i.e., resonant photoproduction
$\gamma p \to \jpsi\,p \to P_c(4450) \to \jpsi\,p$,
as a function of the incident photon laboratory energy $E_\gamma$.

The CM energies of 4.38 and 4.45 GeV correspond to laboratory photon energies
of 9.75 and 10.08 GeV, respectively, well within the capabilities of the
GlueX and CLAS12 detectors at Thomas Jefferson National Accelerator Facility
(JLAB) \cite{Shepherd:2014hsa,Glazier:2010zz}.  

For example, CLAS12 will
produce a tagged photon spectrum via bremsstrahlung from an electron beam,
yielding a total of $5\times 10^7$ photons per second with $6.5 < E_\gamma <
10.5$ GeV and having a spectrum $dN/dE_\gamma = A/E_\gamma$ 
\cite{BattaglieriPC}.  Demanding that
the integral of $dN_\gamma /dE_\gamma$ from 6.5 to 10.5 GeV be $5\times 10^7$
photons per second, we find
\beq
dN_\gamma/dE_\gamma = 1.0 \times 10^8~{\rm photons/s}/E_\gamma~.
\label{dNgammadEgamma}
\eeq
This spectrum may be used to estimate the signal [using Eq.\ (\ref{eqn:jpsip})]
and background for the resonances $X_A$ and $X_B$ with arbitrary spin.

As a sample calculation of the expected number of events we consider here
resonant production of $P_c(4450)\equiv X_B$.  The CM energy range 
$m_B{-}\Gamma_B/2 < E < m_B{+}\Gamma_B/2$
corresponds to $9.99\ {\rm GeV} < E_\gamma < 10.18$ GeV, i.e., $\Delta E_\gamma
=0.19$ GeV.  From Eq.~\eqref{dNgammadEgamma} we then obtain the number of
photons corresponding to $E$ under the resonance peak:
\beq
N_\gamma =
\int\displaylimits_{9.99\,{\rm GeV}}^{10.18\,{\rm GeV}} 
{dN_\gamma\over dE_\gamma} dE_\gamma \approx 2 \times 10^6~{\rm
photons/s}~.
\label{photon_flux}
\eeq

Since the photon beam intensity is given in terms of number of photons per
second, rather than in the usual units of luminosity, we shall use here the
GlueX rule of thumb that an intensity of $10^5$ $\gamma$/s will produce about
$10^4$ events per 1 $\mu$b of cross section per day of running
\cite{Shepherd:2014hsa}.
So with a peak cross section of 1.35 \,$(B_{\rm out})^2 \mu$b, a branching
fraction ${\cal B}(J/\psi \to e^+ e^-) = (5.971 \pm 0.032)\%$ \cite{PDG}
and the photon flux \eqref{photon_flux} we should expect $1.6 \times 10^4\,
(B_{\rm out})^2$ events per day of running.  While this looks large, we do not
know the magnitude of $B_{\rm out}$.

In the region of interest the $E_\gamma$ resolution is 20-30 MeV, corresponding
to 4-6 MeV resolution in $E$. This is much less than the 39 MeV width
(in $E$)  of the $P_c(4450)$ resonance, so it should be possible to resolve the
peak in Fig.~\ref{fig:sigma_photo}.  It is likely that in the future the
$E_\gamma$ resolution will be even better \cite{BattaglieriPC}.
For details of a specific CLAS12 proposal to study $J/\psi$ production with a
tagged polarized photon beam of energy 11 GeV, see Ref.\ \cite{JLAB:PAC39}.
Such a beam enables useful measurements of resonance spin-parity via angular
distributions of the final $e^+e^-$ pair in $J/\psi$ decay \cite{StonePC}.

In the future the GlueX detector \cite{Shepherd:2014hsa,JonesPC} will
complement the reach of CLAS12.  A specific proposal to study $\gamma p \to
J/\psi p$ with $8.7 < E_\gamma < 11.5$ GeV \cite{Seth}, optimized for a peak
in the photon spectrum at 10 GeV \cite{SethPC}, leads one to expect about
$6 \times 10^{-4}$ events/MeV/s/$\mu$b (with $J/\psi \to e^+e^-$).
Integration with respect to CM energy $E$ over a Breit-Wigner resonance with
maximum $\sigma_{\rm peak}$ and width $\Gamma$ multiplies $\sigma_{\rm peak}$
by a factor of $\pi \Gamma/2 = 61.26$ MeV.  But we want to integrate with 
respect to laboratory photon energy $E_\gamma$, so we have to multiply by
$dE_\gamma/dE = E/m_p = 4.743$, giving a factor of 290.5 MeV.
Multiplying by $\sigma_{\rm peak} = 1.35~\mu{\rm b}(B_{\rm out})^2$
one estimates a rate of about $2 \times 10^4(B_{\rm out})^2$ events of the
4450 MeV state per day, roughly consistent with our estimate for CLAS12.
As for energy resolution, GlueX expects a RMS tagged
photon uncertainty around 6 MeV. On a proton target, this translates
into an uncertainty in $E$ of 1 MeV for a 10 GeV photon \cite{JonesPC},
which should enable an accurate scan of the resonance lineshape.

The branching fraction $B_{\rm out}$ cannot be too small, as the $P_c(4450) \to
J/\psi~p$ signal is 4.1\% of the $J/\psi~p$ final state in $\Lambda_b \to K^-
J/\psi~p$ \cite{LHCb}.  If $B_{\rm out}$ is too small, the value of
${\cal B}(\Lambda_b\to K^- P_c)$, with $P_c$ decaying to final states other
than $J/\psi~p$, becomes unreasonably large in comparison with ${\cal B}
(\Lambda_b \to K^- J/\psi~p) = 3 \times 10^{-4}$ \cite{StonePC}.

\subsubsection*{\boldmath%
Comparison with \jpsi\ photoproduction data near threshold}
The elastic \jpsi\ photoproduction cross section for 
$10 < E_\gamma < 13$  GeV has been measured by SLAC and Cornell teams in
1975 and is quite small, below 1~nb
\cite{Camerini:1975cy,Gittelman:1975ix,Brodsky:2000zc,Chudakov2008,%
Aid:1996dn}.
This raises an obvious question:  Why wasn't the $P_c(4450)$ resonance
observed by these experiments? There are several effects, all 
working in the same direction, as listed below. 
\begin{itemize}
\item[a)]
Smearing by poor energy resolution:
The $P_c(4450)$ width is quite small, 39 MeV, corresponding to 
$180$ MeV in terms of photon energy. The photon energy in the early
experiments had a rather large spread. For example, in the Cornell 
study \cite{Gittelman:1975ix} the photon energy was divided into three
intervals: 9.3--10.4, 10.4--11.1, 11.1--11.8 GeV. The narrow peak is
smeared out when convoluted with such a wide energy distribution.
\item[b)]
Mostly forward scattering:
Experiments \cite{Camerini:1975cy,Gittelman:1975ix}
focused on the forward cross section, which is mostly due
to diffractive scattering, while resonance scattering tends to be 
much more isotropic. Therefore only a small fraction of the 
resonant cross section is in the forward direction.
\item[c)]
$B_{\rm out} \ll 1$:
The branching fraction of the resonance into $\jpsi~p$ might be
significantly less than~1.
In this context it is interesting to point out that Ref.~\cite{Camerini:1975cy} 
used vector dominance to derive the estimate
\beq
\left.{d \sigma(\gamma p \to \jpsi\,p)\over dt}\right\vert_{t=0}
\simeq 25~\mu{\rm b}/{\rm GeV}^2\,.
\eeq
Assuming that the forward $\jpsi\,p$ scattering amplitude is
purely imaginary, they then used the optical theorem to derive the bound
$\sigma_{tot}(\jpsi\,p)\leq 0.8$ mb.
\end{itemize}

\section{\boldmath The reaction $\gamma p \to X \to \Upsilon p$}

It was suggested in Ref.\ \cite{mol} that an exotic doubly-heavy meson or
baryon resonance should exist near any threshold if pion exchange is allowed
between the two constituent hadrons.  In particular, there should exist a
relatively narrow $J^P=3/2^-$ resonance near $\Sigma_b B^*$ threshold, or
11.14 GeV, decaying to $\Upsilon(nS) p$.  We shall estimate the cross
section for photoproduction of such a resonance, denoted by $P_b(11140)$.  The
corresponding photon energy in the laboratory is $E_\gamma = 65.66$ GeV.  In
principle such an energy could be achieved using tagged photons from HERA. 

The calculation for resonant $\Y1S\, p$ photoproduction is entirely analogous
to the one for $\jpsi$.  Using the experimental value \cite{PDG} 
$\Gamma(\Upsilon(1S)\to e^+e^-)=1.34$ keV, we obtain
\beq
f_{\Y1S} = 238~{\rm MeV}~.
\eeq
We then find, for a $\Y1S\,p$ resonance of mass $E_R=11.14$ GeV, with
$k^R_{\rm in,out} = (5.530,1.287)$ GeV the (incoming, outgoing) CM 3-momentum
for $E=E_R$, 
\beq
B_{\rm in}/B_{\rm out} = (e f_\Upsilon/M_\Upsilon)^2 f_0
(k_{\rm in}/k_{\rm out}) = (5.82 \times 10^{-5})(0.663)(4.30) = 1.66 \times
10^{-4}~.
\eeq
The photoproduction cross section for such a resonance with width $\Gamma_{\rm
tot}$ is given by 
\beq \label{eqn:Upsilon:p}
\sigma_{BW}(E) = \frac{C_R (B_{\rm out})^2 (k^R_{\rm in}/k_{\rm in})^2
(\Gamma_{\rm tot}^2/4)} {(E-E_R)^2 + (\Gamma_{\rm tot}^2/4)}~,
\eeq
where, for $J=3/2$,
\beq
C_R \equiv \frac{4\pi}{(k^R_{\rm in})^2} \, \
\frac{B_{\rm in}}{B_{\rm out}}= 26.6~{\rm nb}\,.
\eeq

\begin{figure}[t]
\begin{center}
\includegraphics[width=0.96\textwidth]{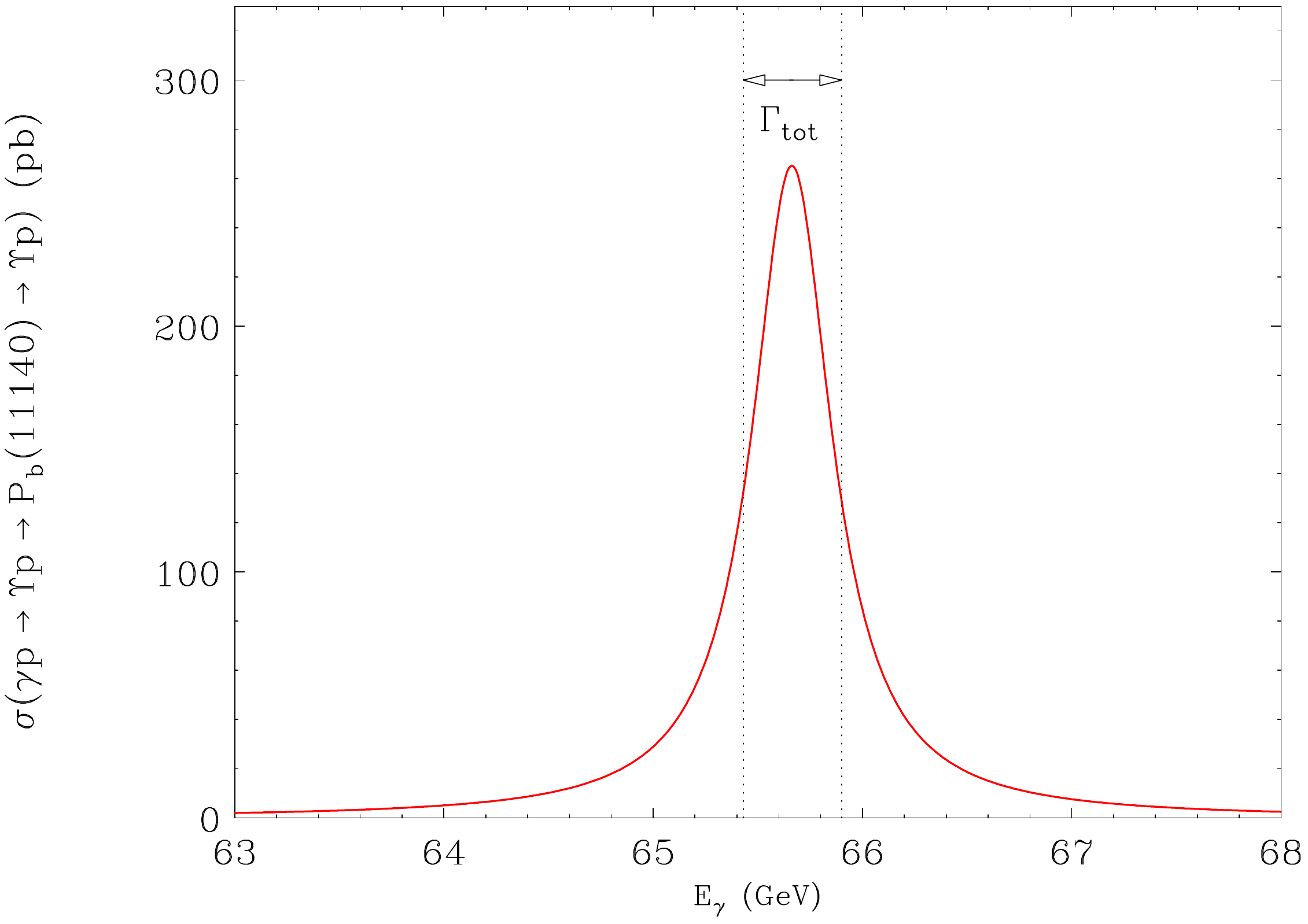}
\end{center}
\strut\vskip-1.2cm
\caption{Cross section for resonant photoproduction $\gamma p \to \Y1S p \to
P_b(11140) \to \Y1S p$ assuming $B_{\rm out} = 0.1$, plotted as function
of the incident photon energy $E_\gamma$. The vertical dotted lines
indicate the width of the $P_b(11140)$ resonance.}
\label{fig:sigma_upsilon_photo}
\end{figure}

The cross section for resonant photoproduction $\gamma p \to \Y1S p \to
P_b(11140) \to \Y1S p$ is shown in Fig.~\ref{fig:sigma_upsilon_photo}
as a function of the incident photon energy $E_\gamma$.  Here we have assumed
the same width as $P_c(4450)$, i.e., $\Gamma=39$ MeV.  The actual width is
likely to be narrow, but its precise value is unknown.  It is given by the
product of the square of the matrix element and the phase space.  Under the
assumption that $B_{\rm out}$ is close to 1, the phase space scales as
$k^R_{\rm out}$ for an S-wave decay. If the matrix element remained unchanged,
it would yield $\Gamma(P_b) = \Gamma(P_c) (k^{P_b}_{\rm out}/k^{P_c}_{\rm out})
= 61$ MeV.  The matrix element is given by the overlap
of the $\Upsilon~p$ and $\Sigma_b B^*$ molecule wave functions. 
This overlap is likely to be less than the overlap of that between the
$\jpsi~p$ and $\Sigma_c \bar D^*$ wave functions as a result of the more
compact nature of the $\Upsilon$, but we do not have a quantitative
estimate.  In the more likely case that $B_{\rm out}$ is much less than 1,
$\Gamma(P_b)$ will depend on details of the molecular binding of $\Sigma_b$
and $B^*$.

The corresponding background is the diffractive process $\gamma p \to
\Upsilon(nS) p$.  A few events of $\gamma p \to \Upsilon(1S) p$ were seen in
1995--7 data by the ZEUS Collaboration at HERA \cite{Breitweg:1998ki}.  They
quoted a ratio 
\beq
\sigma_{\rm el}(\gamma p \to \Upsilon(1S) p)/\sigma_{\rm el}(\gamma p \to
 \jpsi\,p) \sim 5 \times 10^{-3}~.
\eeq
At 11 GeV, Ref.\ \cite{Aid:1996dn} estimated $\sigma_{\rm el}(\gamma p \to
\jpsi\,p) \simeq 10$ nb, yielding $\sigma_{\rm el}(\gamma p \to \Upsilon p)
\simeq 50$ pb.  In later ZEUS data with 62$\pm$12 $\Upsilon(1S)$ events
\cite{Chekanov:2009zz}, $\sigma(\gamma p \to \Upsilon p)$ was measured in
various ranges of center-of-mass energy $W$ to be
\begin{center}
\begin{tabular}{c c c }
$160 \pm 51^{+48}_{-21}$ pb & $60 < W < 130$ GeV & Central $W_0 = 100$ GeV \\
$321 \pm 88^{+46}_{-114}$ pb & $130 < W < 220$ GeV & Central $W_0 = 180$ GeV \\
$235 \pm 47^{+30}_{-40}$ pb & $60 < W < 220$ GeV & ~.\\
\end{tabular}
\end{center}
Comparing cross sections at $W_0 = 100$ and 180 GeV, they scale as $W^{1.18}$.
Assuming this dependence to extrapolate to 11 GeV gives a cross section of
12 pb at that energy.

\section{Conclusions}

The discovery in the LHCb experiment of a narrow resonance of mass 4450 MeV
and a broader enhancement at 4380 MeV, both of which decay to $\jpsi\,p$,
suggests that one search for photoproduction of these states on proton
targets using photons of energy near 10 GeV.  Predicted cross sections are
at an encouraging level above diffractive $\gamma p \to \jpsi\,p$ background.
On the basis of proximity to the $\Sigma_b B^*$ threshold, a predicted state
\cite{mol} near 11.14 GeV should be photoproduced with photons of energy $\sim
66$ GeV.The observation of signals in the $\gamma p \to \jpsi p$ channel would
provide important confirmation of the resonant nature of the LHCb states.
The observation of a narrow resonance in the $\gamma p \to \Upsilon p$ channel 
would be a major new discovery and would strongly indicate existence
of yet additional resonances, along the lines advocated in Ref.~\cite{mol}.

\section*{Acknowledgements}
We thank
Marco Battaglieri,
Stan Brodsky,
Richard Jones,
Sergei Kananov,
Uri Karshon,
Kam Seth,
Matt Shepherd,
Tomasz Skwarnicki,
Stepan Stepanyan,
Sheldon Stone, and
Misha Voloshin for
helpful communications.  The work of J.L.R. was supported in part by the U.S.
Department of Energy, Division of High Energy Physics, Grant No.\
DE-FG02-13ER41958, and was performed in part at the Aspen Center for Physics,
which is supported by National Science Foundation grant PHY-1066293.

\end{document}